\documentclass{article}
\usepackage[T1]{fontenc}
\usepackage[utf8]{inputenc}
\usepackage{ismir_arxiv} % Remove the "submission" option for camera-ready version
\usepackage{amsmath,amssymb, cite,url}
\usepackage{color}
\usepackage{booktabs}   % \toprule, \midrule, \bottomrule, \cmidrule
\usepackage{multirow}   % \multirow
\usepackage{graphicx}   % \resizebox
\usepackage{titlesec}
\usepackage{siunitx} % for \SI command
\usepackage[table]{xcolor} 

\definecolor{myblue}{HTML}{E8F0FE}
\definecolor{myyellow}{HTML}{FFF3E0}
\definecolor{myred}{HTML}{D88C9A}      % define once, reuse
\title{CoJEPA: Combining Contrastive Learning and JEPA for Global-Local Music Representations}

\multauthor
  {Gabriel Meseguer-Brocal \hspace{0.5cm} Yuexuan Kong \hspace{0.5cm} Romain Hennequin}
  {Deezer Research, Paris, France\\
  {\tt\small research@deezer.com}}

\begin{document}

\maketitle

\begin{abstract}

Joint-Embedding Predictive Architecture (JEPA) has shown strong performance in learning rich representations through self-supervised prediction in latent space. However, it typically relies on teacher--student architecture with an EMA to stabilise training, and can tend to yield uninformative representations. Contrastive learning is stable to train and produces strong global representations, but remains limited on local tasks by the global nature of its objective. In this work, we combine both into CoJEPA: a single shared backbone jointly trained with a JEPA objective on masked sequence tokens and a contrastive objective on the class token. The contrastive gradient provides stability, removing the need for an EMA teacher entirely, while JEPA enriches the sequence tokens via local predictions that contrastive learning alone cannot provide. Crucially, no extra parameters are added to the backbone: the same model is guided towards richer representations purely through the design of its training signal. CoJEPA takes the best of both worlds, outperforming or matching both individual methods across global and local MIR tasks, with a particularly strong advantage on tonal and harmonic understanding, and without any task-specific architectural changes. CoJEPA shows that combining objectives with complementary inductive biases can substitute for scale, encouraging future work to invest in smarter training objectives over ever-larger models.
\end{abstract}

\section{Introduction}\label{sec:introduction}

% % SELF SUPERVISE LEARNING INTRO
% Self-supervised learning (SSL) methods boom, highlighting benefits and limitations (ismir tutorial)

% Families.
% \begin{itemize}
%     \item joint embeddings
%     \item masking
%     \item jepa 
%     \item equivariants
% \end{itemize}

% Mixing these families has been already explore (waspa paper: two joint embeddings with one being equivariants, others??). 

% SELF SUPERVISE LEARNING INTRO

% narrative: 
% ssl is cool, useful blabla, especially in music!
% it has 3 big families: jea, equivaraint, masked modelling.
% each has its own limitations: jea, not perform well in local tasks, equivariant is limited to one task, masked modelling -> harder to train, requires bigger models, pixel space does not always make sense
% so... lecun proposed jepa, so cool! prediciton in the latent space, wow! but jepa has ema, hard to train, instable blabla, how can we regularize it? 
% Mo et al.
% however, still needs ema, also has never been done in music
% recently, the emergent properties -> we can take advantage of transformer's class token! we can combine jea with jepa
% our paper contribution: no need for ema, stable to train. improve on both contrastive and jepa, youpi

Self-supervised learning (SSL) has emerged as an impactful paradigm in music information retrieval (MIR) in recent years.
It enables the learning of rich audio representations without requiring heavy manual annotations, thereby circumventing common issues seen in supervised learning, such as label ambiguity, annotation cost, and lack of generalization.
Moreover, SSL aligns with how human learns internal musical representations, through exposure, interaction and prediction~\cite{bamberger1991mind,levitin2012does,honing2011illiterate}, as musical training is not necessary for requiring musical knowledge~\cite{krumhansl_cognitive_2001}.

In this paradigm, a backbone is trained via a pretext task where the supervisory signal is derived directly from the data itself, rather than from human annotations.
The backbone aims to produce useful representations that can be adapted to a wide range of downstream tasks with minimal effort, usually via probing, i.e., training a lightweight classifier on top of the frozen backbone representations.
What makes SSL particularly attractive is not only the little requirement of annotations, its usefulness in cleaning and improving annotations~\cite{meseguer2020data} and the reusability of the learned representations, but also the fact that they have managed to surpass supervised approaches on a broad set of downstream tasks~\cite{won2024musicfm, li2024mert}. Yet a central challenge remains: designing better training objectives that guide models toward more rich and diverse features.

As one of the latest family in SSL, \textbf{Joint-Embedding Predictive Architecture (JEPA)} was recently proposed~\cite{lecun2022path, assran2023ijepa}. The core idea of of JEPA is to predict masked regions directly in a learned latent space. 
In practice, JEPA relies on a teacher--student framework in which a teacher encoder produces latent targets while a student network learns to predict them.
The teacher is updated through an exponential moving average (EMA) mechanism to stabilize training and avoid collapse~\cite{assran2023ijepa}.
However, while powerful in learning representations, JEPA is still hard and unstable to train, with extensive hyperparameter search, and prone to collapse~\cite{assran2023ijepa, balestriero2025lejepa}. At the same time, as an easy-to-train and stable SSL paradigm, contrastive learning performs worse on local tasks since the losses are typically applied to capture global variance and invariance~\cite{mccallum2022supervised, wang2020understanding, kong2025multi}.
Moreover, recent analyses of transformer-based contrastive learning models have revealed emergent musical properties in sequence tokens~\cite{kong2025emergent}. Surprisingly, even when contrastive objectives are applied only to a global class token, sequence tokens still encode local musical information useful for frame-level downstream tasks.

% CORE IDEA
Building on these ideas, we propose CoJEPA an efficient and novel method for combining the strengths of JEPA and contrastive learning. To a traditional JEPA, we apply a contrastive loss on the class token of a transformer, as a regularisation to constrain the latent space, \textbf{removing the need for EMA} and thereby stabilizing training. As a result, CoJEPA improves the quality of global and local representations compared to standard JEPA and contrastive pre-training and it outperforms either method alone on downstream tasks, and remains competitive against a much larger model, MusicFM~\cite{won2024musicfm}.

\begin{figure*}[t]
    \centering
    \includegraphics[width=.9\textwidth]{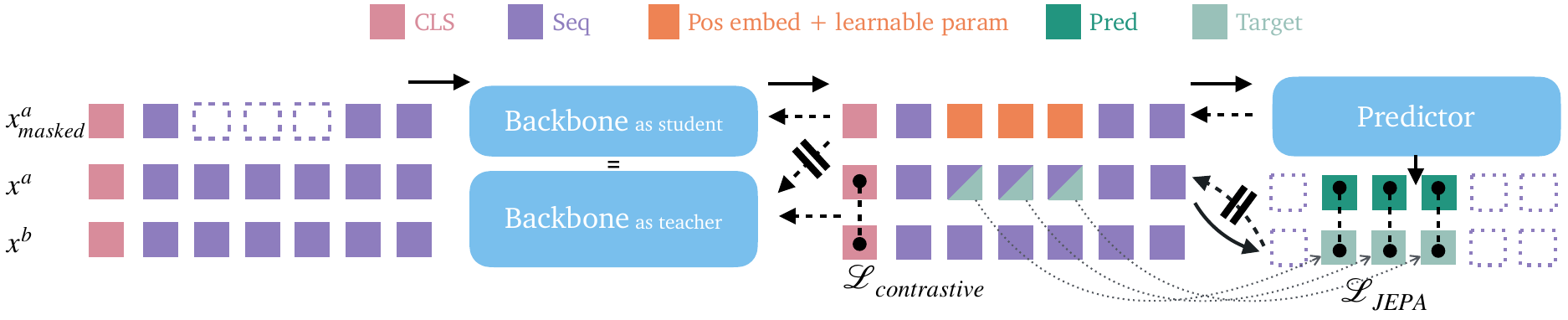}
    \caption{Overview of the proposed method. Two full views $\mathbf{x}_a$ and $\mathbf{x}_b$ are encoded by a shared backbone; their CLS tokens are contrasted via $\mathcal{L}_\text{contrastive}$. The same backbone acts as a teacher (full $\mathbf{x}_a$, stop-gradient) and as a student (masked $\mathbf{x}_a$), a predictor $g_\phi$ reconstructs the missing tokens from context representations and learnable positional placeholders, with gradients from $\mathcal{L}_\text{JEPA}$ flowing back through the predictor and student path. The backbone is thus trained simultaneously as a \textit{teacher} via contrastive and as a \textit{student} via JEPA, removing the need for an EMA teacher.}
    \label{fig:overview}
\end{figure*}

\section{Literature review}\label{sec:sota}
We review prior work on contrastive learning and JEPA: self-supervised methods that propose different ways to derive the training signal from the data itself.

\textbf{Joint embedding architectures (JEA)} learn representations by mapping different views of the same input to nearby locations in a shared embedding space.
These views are typically obtained through data augmentations or different contextual transformations of the same audio excerpt.
Different approaches use different losses to control the invariance and variance needed in the latent space~\cite{chen2020simclr,bardes2022vicreg,zbontar_barlow_2021}.
% Later methods introduced explicit regularization strategies to prevent representational collapse without requiring negative samples, such as variance and covariance regularization~\cite{bardes2022vicreg}.
In music representation learning, joint embedding methods-and more specifically, contrastive learning—have proven particularly effective for learning global semantic representations that transfer well across downstream tasks~\cite{mccallum2022supervised,spijkervet2021contrastive, meseguer2024experimental,ferraro2021enriched,kong2025emergent}.
However, since the losses are applied to a global representation, these approaches often underperform on tasks requiring fine-grained temporal information, such as beat tracking or chord estimation~\cite{kong2025emergent}.

% \textbf{Equivariant SSL} learns representations that transform equivariantly under known input transformations~\cite{quinton2022equivariant, riou2023pesto, kong2024stone, kong2025s, gfeller2020spice,cwitkowitz2024toward}.
% Applying a transformation to the input is expected to induce an equivariant transformation in the latent representation. 
% Specific to this family, the pretext task is explicitly aligned with the downstream task of interest. 
% For example, transformations related to pitch shifting or tempo scaling can directly encode pitch or tempo information in the latent space.
% While they can reach comparable even surpass supervised methods for specialized tasks, equivariant approaches are tightly coupled to the transformation being modelled, therefore less transferable across a broader range of downstream tasks.

\textbf{Masking modeling methods} learn by predicting masked portions of the input~\cite{he2022mae, hsu2021hubert}. By forcing the model to predict masked regions from contextual information, these methods encourage representations to capture rich information. In audio and music, masking-based approaches have shown strong performance, with models such as MusicFM~\cite{won2024musicfm} and MERT~\cite{li2024mert} establishing state-of-the-art results across a wide range of MIR tasks. However, reconstructing signals directly in the input domain may encourage the model to focus on low-level details that are not always aligned with perceptually meaningful musical semantic features.

To address the limitations of masked modelling, \textbf{Joint-Embedding Predictive Architectures (JEPA)} was recently proposed~\cite{lecun2022path, assran2023ijepa}. Rather than reconstructing missing content in the raw input space, JEPA predicts masked regions directly in a learned latent space. 
This shifts the objective away from low-level reconstruction and toward semantic prediction.
It has shown strong performance in audio and music domain~\cite{riou_stem-jepa_2024,quelennec2025matpac++}.
% In practice, JEPA relies on a teacher--student framework in which a teacher encoder produces latent targets while a student network learns to predict them.
% The teacher is updated through an exponential moving average (EMA) mechanism to stabilize training and avoid collapse~\cite{assran2023ijepa}.
Although effective, JEPA requires EMA, therefore remains difficult to optimize in practice due to training instability and sensitivity to hyperparameters~\cite{chenexploring,tianunderstanding}.

\textbf{Equivariant SSL} learns representations that transform equivariantly under known input transformations~\cite{quinton2022equivariant, riou2023pesto, kong2024stone, kong2025s, gfeller2020spice,cwitkowitz2024toward}.
Applying a transformation to the input is expected to induce an equivariant transformation in the latent representation. 
Specific to this family, the pretext task is explicitly aligned with the downstream task of interest.
While effective for specialized tasks, equivariant approaches are tightly 
coupled to the modelled transformation and therefore less transferable 
across diverse downstream tasks.

While these works have traditionally been studied in isolation, recent work has begun exploring hybrid approaches that leverage multiple SSL principles simultaneously.
For instance, in MIR, \cite{kong2025multi} combines joint embedding learning with equivariance constraints, demonstrating that incorporating geometric structure can enhance representation quality.
Similarly, other works have explored combinations such as contrastive learning with masked prediction~\cite{gong2023cavmae, chung2021w2vbert}, suggesting that different SSL objectives capture complementary aspects of the data structure.

Furthermore, to address the limitations of JEPA, recent work combines predictive latent-space learning with regularized joint embedding objectives~\cite{mo2024cjepa}, by combine I-JEPA~\cite{assran2023ijepa} with VICReg~\cite{bardes2022vicreg}, introducing explicit variance and covariance regularization terms to stabilize the latent space and prevent collapse. 
However, EMA is still necessary to stabilize the training.

\section{Methodology}\label{sec:method}

In this section, we describe the two SSL paradigms at the core of our framework, contrastive learning and JEPA, and present how we combine them into a unified training framework.
We work with an updated ViT-1D transformer~\cite{kong2025emergent} (Section~\ref{sec:model}) as the backbone encoder, which processes audio inputs as sequences of tokens, with a class token prepended at the beginning of the token sequence.
This choice is central to our approach, as the self-attention mechanism and the class token architecture of transformers play a key role in enabling the combination of both paradigms.

\subsection{JEPA}
\label{sec:jepa}

JEPA~\cite{lecun2022path, assran2023ijepa} extends masked modeling by operating in the representation space: tokens are masked after encoding, and a predictor is trained to reconstruct the missing latent representations from the visible ones. Given a sequence of $T$ input audio tokens $\mathbf{x} = \{x_1, \dots, x_T\}$ where $x_i \in \mathbb{R}^{d_{\mathrm{e}}}$ (see Section~\ref{subsec:inputs}):

% JEPA has three elements:
% - a teacher that creates the target seen the whole sequence 
% - a studend that process only some elements of the sequence
% - a predictor that takes as input the output of the student, some prior info about the missing tokens predict the missing ones
% - the loss happens between only at the masked tokens between the output of the predictor and the output of the teacher

\textbf{Teacher.} A teacher encoder $f_{{\theta}_t}$ processes the \textit{full} sequence $\mathbf{x}$, producing a sequence of token representations $\mathbf{z}^t = [{z}_1^t, \dots, {z}_T^t] = f_{{\theta}_t}(\mathbf{x})$, where each ${z}_i^t \in \mathbb{R}^{d_{\mathrm{e}}}$ captures the contextualised semantics of the whole sequence. The tokens at the masked positions serve as \textit{target} tokens.

\textbf{Student.} A student encoder $f_{\theta_s}$ processes only the visible (\textit{context}) tokens $\mathbf{x}^c = \{x_i : m_i = 1\}$, defined by a binary mask $\mathbf{m} \in \{0,1\}^T$, producing $\sum_i{m_i}$ \textit{context} tokens $\mathbf{z}^c=\{z_i^c : m_i = 1\}$ with ${z}_i \in \mathbb{R}^{d_{\mathrm{e}}}$ .

\textbf{Predictor.} A lightweight predictor $g_\phi$ takes $\mathbf{z}$ and prior information about the missing tokens as input; commonly, the positional embeddings $p_i$ of the masked tokens. It then estimates the representations:

$$\hat{\mathbf{z}} = g_\phi(\mathbf{z}^c, \{ p_i : m_i=0\})$$%\mathbf{p}_t)

\noindent $g_\phi$ is itself a transformer where masked tokens attend to context positions via cross-attention, encouraging $\mathbf{z}$ to encode useful information for predicting the missing content.

\textbf{Loss.} The training objective combines L2 distance and cosine similarity between predicted and target tokens, computed only at the masked positions:

$$\mathcal{L}_{\text{JEPA}} = \frac{1}{\sum_i{m_i}} %\frac{1}{|\bar{\mathbf{z}}_t|} 
\sum_{i:\, m_i = 0} 
\left( \| \hat{z}_i - z^t_i \|_2^2 + 1 - \frac{\hat{z}^t_i \cdot z^t_i}
{\|\hat{z}^t_i\| \|z^t_i\|} \right)$$

The L2 term penalises magnitude differences and prevents scale ambiguity. Cosine similarity enforces directional alignment between predicted and target. Their combination is motivated by the decomposition of SSL objectives into alignment and uniformity components~\cite{wang2020understanding}: L2 alone is prone to representation collapse through magnitude shrinkage, whereas cosine similarity induces higher-order gradient dynamics that provide stronger collapse resistance in non-contrastive settings~\cite{bao2023feature}.

By predicting in latent space rather than in the raw signal domain, JEPA encourages the model toward semantically meaningful representations, abstracting away low-level reconstruction details.

A key challenge in JEPA is that the target $\mathbf{z}^t$ is not fixed but evolve throughout training, since the teacher encoder $f_{\theta_t}$ is updated through an EMA of the student weights:

$${\theta}_t \leftarrow \tau \theta_t + (1 - \tau)\theta_s$$ 

\noindent where $\tau \in [0, 1)$ is a momentum coefficient progressively increased toward $1$~\cite{grill2020byol, assran2023ijepa}. This moving target creates a co-adaptation dynamic: the student is pushed toward 
semantically rich representations in order to be useful to predict the targets, while the targets are pressured to become easier to predict, leading to the potential representational 
collapse~\cite{mo2024cjepa}. Although EMA slows the process by ensuring targets evolve smoothly, it does not fully prevent collapse and does not necessarily ensure useful features~\cite{balestriero2025lejepa, mo2024cjepa, bardes2022vicreg}, while also introducing more hyperparameters related to $\tau$.

\subsection{Contrastive learning}
\label{sec:jea}
Contrastive learning learns representations by mapping different views of the same input to nearby locations in a shared embedding space.
We obtain the two views by sampling from the same track two random temporally proximate segments~\cite{ferraro2021enriched, mccallum2022supervised,  spijkervet2021contrastive, meseguer2024experimental, kong2025emergent}. 
Intuitively, two excerpts from the same song should share sufficient musical semantics to yield similar representations, while segments from different tracks should be distinguishable. Given a batch of $B$ audio samples, each sample yields two views $\mathbf{x}$ and $\mathbf{x}'$.
A dedicated class token is prepended at the beginning of the token sequence, i.e., position 0, initialized with trainable parameters and the average of the sequence tokens, used to aggregate information from the full sequence via self-attention.
Two views are passed through the same encoder and class tokens are obtained as: $z_0 = f_{\theta}(\mathbf{x})_{\texttt{[CLS]}},
z'_0 = f_{\theta}(\mathbf{x}')_{\texttt{[CLS]}}$.

\textbf{Loss.} 
We use the NT-Xent loss~\cite{chen2020simclr}.
% Given a batch of $N$ samples, each yielding two views, the $2N$ representations are denoted $\{\mathbf{z}_k\}_{k=1}^{2N}$.
For a positive pair $(z_{0, q}, z'_{0, q})$ obtained from the same audio sample q (we add q for a clearer demonstration of contrastive loss), the per-pair loss is defined as:

$$\ell(q) = -\log \frac{\exp\!\left(\text{sim}(z_{0, q},z_{0, q}') / t\right)}
{\displaystyle\sum\limits_{k \neq q}
\exp\!\left(\text{sim}(z_{0, q}, z_{0, k}) / t\right)},$$
the denominator includes representations from all other samples in the batch, which act as negative examples.
The total loss $\mathcal{L}_{\text{NT-Xent}}$ is computed for all positive pairs.

% $$\mathcal{L}_{\text{NT-Xent}} = \frac{1}{2N} \sum_{k=1}^{N} 
% \left[ \ell(2k-1,\, 2k) + \ell(2k,\, 2k-1) \right]$$

% \noindent where $\text{sim}(\mathbf{u}, \mathbf{v}) = \mathbf{u}^\top \mathbf{v} / 
% (\|\mathbf{u}\| \|\mathbf{v}\|)$ is the cosine similarity and $t > 0$ is a 
% temperature hyperparameter.

\subsection{Combining JEPA and contrastive learning}
\label{sec:combining}

Recent work has shown that transformers trained with a contrastive objective on the class token already develop semantically meaningful representations in their sequence tokens, even without any direct supervision on them~\cite{kong2025emergent}. This suggests that contrastive learning naturally sets up a representation space where JEPA can operate effectively and benefit from. We leverage this observation and combine the two paradigms into CoJEPA: contrastive learning provides a well-structured, stable global embedding space, while JEPA explicitly pushes individual sequence tokens to encode rich local information. \figref{fig:overview} provides an overview of the proposed method and the flow of gradients through each objective.

\textbf{Shared backbone.} Our key design contribution is to use a single shared backbone $f_{\theta}$ (i.e. with $\theta={\theta_t}={\theta_s}$) for the teacher and student roles, \textbf{without any EMA mechanism}. Depending on the input it receives, $f_\theta$ operates in two modes (annotations slightly differ from Section~\ref{sec:jepa}):

\begin{itemize}
    \item \textbf{Teacher mode}: $f_\theta$ processes the \textit{full} sequence $\mathbf{x}$ and produces a complete representation $\mathbf{z} = f_\theta(\mathbf{x})$, including a class token $z_0$ and sequence tokens $\mathbf{z}_{1:N}$.
    \item \textbf{Student mode}: $f_\theta$ processes only the visible context tokens $\mathbf{x}^c$, producing context representations $\mathbf{z}^c$.
\end{itemize}

\textbf{Forward pass.} Each training step processes two views $(\mathbf{x}, \mathbf{x}')$ sampled from the same track. Both views are passed through $f_\theta$ in teacher mode, producing:

$$\mathbf{z} = f_\theta(\mathbf{x}), \qquad \mathbf{z}' = f_\theta(\mathbf{x}')$$

The class tokens $z_{0}$ and $z'_{0}$ are used to compute the NT-Xent contrastive loss $\mathcal{L}_{\text{NT-Xent}}$. In parallel, the sequence tokens at the masked positions $\{z_i : m_i = 0\}$ serve as JEPA prediction targets. These are detached from the computational graph ($\text{sg}(\{z_i : m_i = 0\})$), so that the JEPA loss does not propagate through teacher mode. The masked views $\mathbf{x}^c$ are then processed in student mode, and the predictor recovers the target representations:

$$\hat{\mathbf{z}} = g_\phi\!\left(f_\theta(\mathbf{x}^c),\, \{ p_i : m_i=0\}\right)$$

\textbf{Gradient flow.} The backbone $f_\theta$ receives gradients from two sources:

\begin{itemize}
    \item $\mathcal{L}_{\text{NT-Xent}}$: flows through the teacher forward pass, shaping the global structure of the embedding space.
    \item $\mathcal{L}_{\text{JEPA}}$: flows through the predictor and the student forward pass, pushing sequence tokens to encode locally predictive information.
\end{itemize}

\noindent The total training objective is:

$$\mathcal{L} = \mathcal{L}_{\text{NT-Xent}} + \mathcal{L}_{\text{JEPA}}$$

\textbf{Eliminating EMA.} Unlike standard JEPA, CoJEPA does not require EMA mechanism. The contrastive loss directly regularizes the backbone, preventing representational collapse by enforcing a well-structured global embedding space. The teacher is thus not a slowly drifting copy of the student but the same model, updated jointly and stably through $\mathcal{L}_{\text{NT-Xent}}$.

\section{Model}
\label{sec:model}

\subsection{Inputs}
\label{subsec:inputs}
Our model takes as input the raw waveform sampled at $16$~kHz and computes a normalized mel-frequency spectrogram with $128$ frequency bins and a frame rate of $31.5$~Hz. This representation is widely used in MIR for contrastive learning and audio tagging task~\cite{meseguer2024experimental, mccallum2022supervised}.
The spectrogram is patchified such that each time frame is projected into the transformer embedding dimension, yielding a sequence of tokens 
$\boldsymbol{x} = [\boldsymbol{x}_{1}, \dots, \boldsymbol{x}_{T}]$ 
with $\boldsymbol{x}_i \in \mathbb{R}^{d_{\mathrm{e}}}$, $d_{\mathrm{e}} = 192$, 
and $T = 126$ for a $4$-second segment.

\subsection{Backbone}

The backbone is updated based on a Vision Transformer with 1-D patches (ViT-1D)~\cite{kong2025emergent}, using an embedding dimension of $192$, a depth of $12$ blocks, and $3$ attention heads per block, Tiny configuration from the original ViT paper~\cite{dosovitskiy2020image}. 
We incorporate three architectural upgrades from modern large language models~\cite{grattafiori2024llama}: (a) rotary positional embeddings (RoPE) in place of standard learned positional embeddings, (b) RMSNorm instead of LayerNorm, and (c) feed-forward layers with gated activations. These upgrades result in a backbone with 7,1M trainable parameters.

\textbf{Class token.} The class token (CLS token) is a special token prepended to the input sequence, responsible for aggregating global sequence information via self-attention. It is defined as a learnable vector of dimension $d_{\mathrm{e}} = 192$, to which we add the mean of the projected input tokens, $\boldsymbol{x}_\mathrm{spe}$. And it is where the joint-embedding loss is applied.

\textbf{Masking technique.} We apply contiguous temporal masking: a single block of consecutive time frames is masked, with all frequency bins zeroed out. Since our patchification maps one time frame to one token, this is equivalent to masking a contiguous block of tokens in the input sequence. 
The mask length per sequence is sampled uniformly from $50\%$ to $75\%$ of the sequence length $T$.
The starting position is drawn uniformly at random from the valid range. Each sample in the batch receives an independently sampled mask length, introducing variability in the amount of context available to the student encoder.

\subsection{Predictor}
The predictor has the same architecture as the backbone but smaller: an embedding dimension of $192$, a depth of $6$ blocks, resulting in $3.5$M parameters. Ablation of these design choices is left for future work; which is relevant given that intermediate layers of the backbone are the ones capturing the richest representations, suggesting the predictor's depth plays a role in shaping them Section~\ref{sec:exp}.

\textbf{Masked tokens.} At each masked position, a learnable token of dimension $192$ is injected as a placeholder, to which the corresponding rotary embedding is added to encode the position of the missing token within the sequence. This allows the predictor to attend to masked positions while distinguishing them from the visible context tokens.

\begin{table}[t]
\centering
\resizebox{\columnwidth}{!}{%
\begin{tabular}{llrr}
\toprule
\textbf{Task} & \textbf{Dataset} & \textbf{Size} & \textbf{nb. classes} \\
\midrule
\rowcolor{myblue}
\multicolumn{4}{l}{\textit{Global tasks}} \\[2pt]
% \midrule
Tagging     & MTAT                      & 25k & 50 \\
Key         & FMAKv2                    & 5.5k & 24 \\
Pitch       & TinySOL                   & 3k & 84 \\
Instruments & TinySOL                   & 3k & 14 \\
BPM         & GiantSteps\textsubscript{tempo}                 & 664 & 72 \\

% \midrule
% \midrule
\rowcolor{myyellow}
\multicolumn{4}{l}{\textit{Local tasks}} \\[2pt]
% \midrule
Chords      & RWC\textsubscript{pop} + SWD             & 124 & 25 \\
Beat        & Ballroom + GTZAN\textsubscript{rhythm}   & 1.7k & -- \\
\bottomrule
\end{tabular}}
\caption{Statistics of downstream datasets. We omit the nb. classes for beat tracking (More details in~\ref{subsec:downstream}).}
\label{tab:datasets}
\end{table}

% ── GLOBAL TASKS ─────────────────────────────────────────────────────────────
\begin{table*}[ht]
\centering
\resizebox{\textwidth}{!}{%
\begin{tabular}{ll ccccc | ccccc | ccccc | c}
\toprule
& & \multicolumn{5}{c}{\textit{Layer 4 - early}}
  & \multicolumn{5}{c}{\textit{Layer 8 - mid}}
  & \multicolumn{5}{c}{\textit{Layer 12 - last}}
  & \multicolumn{1}{c}{\textit{Baseline}} \\
\cmidrule(lr){3-7}\cmidrule(lr){8-12}\cmidrule(lr){13-17}\cmidrule(l){18-18}
& & JEPA
  & \multicolumn{2}{c}{Contrastive}
  & \multicolumn{2}{c}{CoJEPA}
  & JEPA
  & \multicolumn{2}{c}{Contrastive}
  & \multicolumn{2}{c}{CoJEPA}
  & JEPA
  & \multicolumn{2}{c}{Contrastive}
  & \multicolumn{2}{c}{CoJEPA}
  & MusicFM \\
\cmidrule(lr){3-7}\cmidrule(lr){8-12}\cmidrule(lr){13-17}\cmidrule(l){18-18}
\textbf{Task} & \textbf{Metric}
  & seq & CLS & seq & CLS & seq
  & seq & CLS & seq & CLS & seq
  & seq & CLS & seq & CLS & seq & seq \\
  
\midrule
\multirow{2}{*}{Tagging}
  & MAP\textsubscript{macro}     & \cellcolor{myred!40}{.404} & .265 & .266 & .410 & .429 & .393 & .266 & .266 & .436 & \cellcolor{myblue}\textbf{.436} & .376 & .393 & \cellcolor{myyellow}{.395} & .395 & .421 & \underline{.454} \\
  & ROC-AUC\textsubscript{macro} & \cellcolor{myred!40}{.888} & .804 & .803 & .891 & .896 & .884 & .804 & .802 & \cellcolor{myblue}\textbf{.903} & .902 & .875 & \cellcolor{myyellow}{.882} & .881 & .884 & .897 & \underline{.906} \\

\midrule
Key & w-acc & \cellcolor{myred!40}{.245} & .232 & .243 & .633 & .609 & .167 & .234 & .236 & \underline{\cellcolor{myblue}\textbf{.640}} & .518 & .147 & .539 & \cellcolor{myyellow}{.550} & .503 & .396 & .558 \\

\midrule
Pitch
  & acc & \cellcolor{myred!40}{.959} & .962 & \cellcolor{myyellow}{.966} & \underline{\cellcolor{myblue}\textbf{.983}} & .979 & .877 & .962 & .962 & .973 & .973 & .699 & .925 & .962 & .873 & .921 & .969 \\

\midrule
Instruments
  & acc & \cellcolor{myred!40}{.938} & .692 & .688 & .935 & .945 & .911 & .695 & .688 & .949 & \underline{\cellcolor{myblue}\textbf{.952}} & .873 & .942 & \underline{\cellcolor{myyellow}\textbf{.952}} & .942 & .925 & .897 \\  

\midrule
\multirow{2}{*}{BPM}
  & acc@1 & .375 & .444 & .403 & .389 & .604 & .806 & .389 & .396 & .757 & .826  & \cellcolor{myred!40}{.833} & \underline{\cellcolor{myyellow}{\textbf{.854}}} & .847 & \cellcolor{myblue}{.840}  & .826 & .771 \\
  & acc@2 & .389 & .444 & .403 & .389 & .632 & .847 & .389 & .396 & .799 & .861  & \cellcolor{myred!40}{.875} & \underline{\cellcolor{myyellow}{\textbf{.917}}} & .895 & \cellcolor{myblue}{.910}  & .875 & .819 \\

\bottomrule
\end{tabular}}
\caption{Global tasks at three transformer depths. JEPA is evaluated only on mean-pooled sequence tokens (seq). Contrastive and CoJEPA are evaluated on both the CLS token and mean-pooled sequence tokens. MusicFM is reported as a single baseline without layer ablation. \underline{Underlined}: best overall. \textbf{Bold}: best among our three models. \colorbox{myred!40}{Red}: best JEPA. \colorbox{myyellow}{Yellow}: best Contrastive. \colorbox{myblue}{Blue}: best CoJEPA. }
\label{tab:global}
\end{table*}

\section{Experiments}
\label{sec:exp}

\subsection{Pretraining}
We train three models: JEPA, contrastive, and CoJEPA. For the JEPA baseline we follow the standard EMA-based training procedure~\cite{assran2023ijepa}.

\textbf{Dataset.} We use an in-house dataset consisting of about $2$M 
full tracks to pre-train our models. The dataset is curated to be as musically 
balanced and diverse as possible, in order to improve generalization to unseen data.

\textbf{Hyperparameters.} All models are trained for $999$ epochs with $512$ 
steps per epoch and a batch size of $128$. 
We note that the batch size of $128$ is small relative to common contrastive learning practice, which typically relies on large batches to provide sufficient negative pairs~\cite{mccallum2022supervised}. 
We use the AdamW optimizer with $\beta = (0.9, 0.98)$ and a cosine learning rate schedule with $10$ warm-up epochs and a minimum learning rate of $1.5 \times 10^{-6}$. Weight decay follows a cosine schedule from $0.04$ to $0.01$. For the JEPA EMA baseline, the teacher momentum is scheduled from $0.996$ to $0.9999$ over the full training. The masking ratio is fixed at $75\%$.
All models are trained on a single GPU. 

\subsection{Downstream tasks}
\label{subsec:downstream}

We group tasks into \textbf{global} (sequence-level prediction: tagging, key, pitch, BPM) and \textbf{local} (frame-level prediction: beat tracking, chord estimation). Table~\ref{tab:datasets} summarizes all evaluation datasets.
Global tasks require a sequence-level prediction capturing the overall musical content, while local tasks require frame-level predictions at a resolution adapted to the downstream task.

\textbf{Global tasks.}
For music tagging, we train and evaluate on MagnaTagATune using the split of~\cite{lee2017sample}, and report macro ROC-AUC and mAP. 
For instrument recognition and pitch estimation, we use TinySOL~\cite{cella2020orchideasol} with a random 8:1:1 split, evaluating top-1 accuracy over 14 instrument classes and 82 pitch classes respectively.
For key estimation, we train on FMAKv2~\cite{kong2024stone} and test on GiantSteps~\cite{knees2022giansteps}, using the MIREX weighted accuracy metric~\cite{raffel2014mir_eval}. 
For BPM estimation, we use Giantsteps-tempo-dataset\cite{knees2022giansteps} with a split of 8:1:1 for training, validation and evaluation. 
We report $\mathrm{ACC}_1$, accuracy within a 4\% tolerance, and $\mathrm{ACC}_2$ accuracy with the same tolerance additionally allowing octave errors of 2, 3, 1/2, and 1/3.

\textbf{Local tasks.}
For beat tracking, we train on Ballroom Dataset~\cite{gouyon2004ballroom} and evaluate on GTZAN Rhythm~\cite{marchand2015gtzanrhythm}.
We apply temporal smoothing, reporting the $F_1$-score with a $\pm$\SI{70}{\milli\second} tolerance window.
For chord estimation, we use a combined corpus of RWC-POP~\cite{goto2002rwc} and Schubert Winterreise Dataset~\cite{weiss2021schubert} with an 8:1:1 split, and evaluate frame-level accuracy over 24 major/minor chord classes plus a ``no chord'' class.

% ── LOCAL TASKS ──────────────────────────────────────────────────────────────
\begin{table}[t]
\centering

\resizebox{.8\columnwidth}{!}{%

\begin{tabular}{ll cc ccc}
\toprule
& & \multicolumn{1}{c}{\textit{Chords}} & \multicolumn{2}{c}{\textit{Beat}} \\
\cmidrule(lr){3-3}\cmidrule(lr){4-5}
\textbf{Layer} & \textbf{Model} & overlap & F-measure & P-score \\
\midrule
\multirow{3}{*}{4 - early}
  & JEPA       & \cellcolor{myred!40}{.211} & .778 & .745 \\
  & Cont.      & \cellcolor{myyellow}{.258} & .549 & .551 \\
  & CoJEPA     & \cellcolor{myblue}\textbf{.399} & .778 & .747 \\
\midrule
\multirow{3}{*}{8 - mid}
  & JEPA       & .097 & .803  & .774 \\
  & Cont.      & .256 & .551 & .551 \\
  & CoJEPA     & .275 & \cellcolor{myblue}{.793} & \cellcolor{myblue}{.764} \\

\midrule
\multirow{3}{*}{12 - last}
  & JEPA       & .070 & \cellcolor{myred!40}\textbf{.806 }& \cellcolor{myred!40}\textbf{.780} \\
  & Cont.      & .241 & \cellcolor{myyellow}{.737} & \cellcolor{myyellow}{.707} \\
  & CoJEPA     & .160 & .789 & .757 \\

\midrule
\midrule
\textit{Baseline}  & MusicFM       & .193 & \underline{.866} & \underline{.851} \\

\bottomrule
\end{tabular}}
\caption{Local tasks. The probe is applied independently to each sequence token (no pooling).}
\label{tab:local}
\end{table}

\subsection{Evaluation protocols}

We follow the standard SSL linear probing protocol: backbone weights are frozen and a single linear layer ($192 \to $ nb\_{classes} + softmax) is trained per task for $50$~epochs, $128$~steps per epoch, batch size~$128$, learning rate $10^{-3}$ and fixed random seed. We probe at three transformer depths: layer~$4$ (early), layer~$8$ (mid), and layer~$12$ (last), since different layers capture qualitatively different information~\cite{pasad2021layer, kong2025emergent}, allowing us to better attribute performance differences to each training objective.

\textbf{Global tasks.} We probe a single fixed-length sequence representation 
using one of two pooling strategies:

\begin{itemize}
    \item \textbf{CLS token}: for Contrastive and CoJEPA, explicitly optimized to aggregate global information. 
    \item \textbf{Mean-pooled sequence tokens}: The only option for JEPA (no CLS token). We computed for all three models as it provides a complementary view of how well the local tokens capture global musical content.
\end{itemize}

\textbf{Local tasks.} The linear probe is applied independently to each token in the sequence.
% \textcolor{red}{TODO: talk about independent head??}
This evaluates the quality of token-level representations and is relevant for assessing the benefit of the JEPA objective on local features.

\textbf{Baseline.} We compare against MusicFM~\cite{won2024musicfm}, a recent SSL foundation model based on a Conformer encoder ($330$M parameters, $d_e=1024$). We extract features from its last transformer layer only and train our linear probes on $10$-second segments (consistent with pretraining) or on the longest available segment when the audio is shorter. We acknowledge that intermediate layers may yield stronger results~\cite{pasad2021layer}. MusicFM thus serves as an indicative reference point against a recent large-scale model.

\section{Results}\label{sec:results}

\textbf{Comparison of JEPA, Contrastive, and CoJEPA.}
Tables~\ref{tab:global} and~\ref{tab:local} report linear probing results across all tasks at three layers. CoJEPA achieves the best or tied-best result on the majority of tasks, confirming that combining both objectives takes the best of both worlds: the local representational richness of JEPA and the globally structured embedding space of contrastive learning. The only exceptions are rhythmic tasks: BPM, where Contrastive (CLS) marginally leads, and beat tracking, where JEPA leads; yet CoJEPA remains competitive in both cases. Tasks requiring different musical understanding: tagging, key, and chords; show the largest benefit, with substantial margins over either individual method.

\textbf{JEPA representations peak early.} For most tasks, JEPA representations are strongest at layer~$4$ and degrade with depth: pitch drops from $.959$ to $.699$, key from $.245$ to $.147$, and chords from $.211$ to $.070$. This suggests that the JEPA pretext task, as a local temporal prediction objective, produces its most transferable features early in the network but does not provide sufficient guidance for deeper layers to develop music rich semantic representations. This is consistent with findings showing that JEPA alone requires large-capacity models to avoid representation collapse in later layers~\cite{assran2023ijepa}. The notable exception is BPM and beat tracking, where JEPA improves with depth (BPM acc@1: $.375 \to .833$, beat F-measure: $.778 \to .806$). This suggests that rhythmic regularity, which is temporally localised and periodically structured, is naturally captured by the JEPA when representations mature, unlike tonal and semantic features that require explicit auxiliary objectives to steer training beyond JEPA's inherent temporal bias. 

\textbf{Contrastive representations crystallise at the last layer.} Contrastive models produce weak early-layer representations for most tasks (e.g., key w-acc at layer~$4$: $.243$, instruments at layer~$4$: $.692$) but improve dramatically at the final layer (key: $.550$, instruments: $.952$). This indicates that contrastive learning encodes discriminative information primarily in the last-layer global embedding, leaving intermediate representations relatively uninformative. On local tasks, this weakness is most apparent: beat F-measure reaches only $.737$ even at the last layer, well below JEPA ($.806$) and CoJEPA ($.793$), reflecting the lack of direct supervision on individual sequence tokens. 

\textbf{CoJEPA peaks at intermediate depth.} CoJEPA achieves its best performance at layer~$8$ for most tasks: tagging MAP ($.436$), key ($.640$), instruments ($.952$), beat ($.793$); a qualitatively distinct behaviour from JEPA (peaks early) and Contrastive (peaks late), suggesting the two objectives interact to produce well-structured representations at intermediate depth. BPM follows the rhythmic trend of JEPA, improving steadily with depth (CLS at L12: $.840$). Particularly striking is the early emergence of tonal representations. Pitch, and chords peak at layer~$4$ ($.983$ and $.399$) and key at layer-8 ($.64$). This is unexpected given that harmonic information has been shown to be difficult to capture in general music representation learning models~\cite{li2024mert, ding2024parameter} without a specifically designed training objective~\cite{kong2024stone}. These features are not preserved at depth, likely discarded as contrastive increasingly dominates.

\textbf{CLS vs.\ Sequence Tokens.} For Contrastive models, the gap between CLS and mean-pooled sequence tokens is consistently small, suggesting the contrastive signal passively enriches sequence tokens through self-attention. For CoJEPA, the gap is larger and task-dependent: the CLS token dominates on key ($.640$ vs.\ $.518$ at layer~$8$), While sequence tokens are better on BPM and tagging, particularly at early and intermediate layers where the JEPA objective is most influential. This layer-dependent divergence suggests that the two token types carry complementary information in CoJEPA: the CLS encodes a globally summary, enforced by the contrastive objective; the sequence tokens retain local structure, encouraged by the JEPA objective.

\textbf{Comparison with MusicFM.} Despite its much larger scale, MusicFM leads only on tagging and beat tracking, and matching results for on pitch. As a purely masked prediction model, MusicFM also mirrors the behaviour of JEPA on rhythmic versus tonal tasks (strong on beat, weaker on chords) suggesting that the masked prediction objective, without a global discriminative component, faces the same structural limitations regardless of scale. Given the large gap in model capacity ($7.1$M, $d_{e}=192$ vs $330$M, $d_{e}=1024$), these results further support the effectiveness of principled objective design over sheer scaling.

\section{Conclusion}\label{sec:conclusion}

We presented CoJEPA, a self-supervised framework that unifies JEPA and contrastive learning. The two objectives are complementary: JEPA drives sequence tokens to encode rich local structure; while contrastive learning shapes a globally discriminative embedding space, its gradient preventing collapse without the need for an EMA teacher.
Crucially, both objectives operate on the \textit{same} backbone: rather than adding parameters, CoJEPA guides the model towards richer representations purely through the design of its training signal. Despite its small scale ($7.1$M), CoJEPA outperforms or matches both individual methods across global and local tasks, with a particularly strong advantage on harmonic understanding. Interestingly, the richest representations emerge at intermediate depth rather than the final layer, a behaviour that neither JEPA nor contrastive learning exhibits alone. While CoJEPA is expected to benefit from scaling model and batch sizes (as contrastive learning improves with more negative pairs~\cite{mccallum2022supervised}) its strong performance at small scale shows the benefits of complementary training objectives. We encourage future work to invest more in smarter designs, notably the combination of different self-supervised learning objectives.

% guiding the same model towards a better solution without increasing its capacity. 

\newpage
% For BibTeX users:
\bibliography{jepa}

\end{document}